\documentclass[namedreferences]{solarphysics}
\usepackage[hyperref,optionalrh,solaromanenum]{spr-sola-addons} 
\usepackage{graphicx}                    
\usepackage{color}                       
\usepackage{breakurl}                         

\begin{document}
\sloppy
\begin{article}

\begin{opening}

\title{Polarity imbalance of the photospheric magnetic field}

%
\author[addressref={aff1},corref,email={helena@ev13934.spb.edu}]{\inits{E.S.}\fnm{Elena}\lnm{Vernova}}
\author[addressref=aff1]{\inits{M.I.}\fnm{Marta}~\lnm{Tyasto}}\sep
\author[addressref=aff2]{\inits{D.G.}\fnm{Dmitrii}~\lnm{Baranov}}\sep
\author[addressref=aff1]{\inits{O.A.}\fnm{Olga}~\lnm{Danilova}}\sep

%
\runningauthor{E.S.~Vernova \textit{et al.}}
\runningtitle{Polarity imbalance of the photospheric magnetic field}

\address[id={aff1}]{IZMIRAN, SPb. Filial, Laboratory of Magnetospheric Disturbances, St. Petersburg, Russian Federation}
\address[id={aff2}]{Ioffe Physical-Technical Institute, St. Petersburg, Russian Federation}
\begin{abstract}
Polarity imbalance of the photospheric magnetic field was studied
using synoptic maps of NSO Kitt Peak (1976 -- 2016). Imbalance of
positive and negative fluxes was considered for the fields with
strength $B>50$\,G in the sunspot zone ($5^\circ-40^\circ$) and
for the fields with strength $B<50$\,G at higher latitudes
($40^\circ-90^\circ$). The 22-year periodicity in the imbalance of
positive and negative fields was found which maintained itself
during four solar cycles. While for the sunspot zone the sign of
the imbalance always coincides with the northern hemisphere
polarity, for the high latitudes the sign of the imbalance always
coincides with the southern hemisphere polarity. Good
correspondence of the flux imbalance with the quadrupole moment
$(g_{20})$ of the potential-field source-surface (PFSS) model was
observed. The polarity imbalance of the sunspot zone correlates,
on one hand, with the asymmetry of the magnetic field of the
Sun-as-a-star and, on the other hand, with the sector structure of
the interplanetary magnetic field. The obtained results show the
close connection of the magnetic fields in active regions with the
Sun's polar magnetic field. The weakest fields $B<5$\,G represent
quite a special group with the magnetic flux developing in
antiphase to the fluxes of the stronger fields.
\end{abstract}

%
\keywords{Magnetic fields, Photosphere; Polarity imbalance,
Sunspot zone, Polar field}

\end{opening}

%
\section{Introduction}\label{intro}
Magnetic field of the Sun varies with a 22-year periodicity that
manifests itself  both  in the change of signs of leading and
following sunspots during solar minimum (the Hale's law), and in
the change of  sign of the polar field (polar field reversal)
during period of maximum solar activity (see, \textit{e.g.},
\opencite{charbonneau10}). The distributions of the local and the global
magnetic fields exhibit antisymmetry of polarities with respect to
the solar equator. At the same time, there exists some asymmetry
of  the magnetic fields which is reflected in the asymmetric
distribution  of different forms of solar activity. The
north-south asymmetry was discovered in various manifestations of
solar activity, such as sunspots, flares, or sudden disappearances
of solar prominences (see, \textit{e.g.}, \opencite{carb}; \opencite{bal};
\opencite{swi}; \opencite{deng}, and references therein).

The magnetic fluxes of
the Sun and their imbalance were studied on the basis of different
data that characterize the magnetic activity. Asymmetry of the
leading and following sunspot polarities in an active region was
studied by many authors (see, \textit{e.g.}, \opencite{van}; \opencite{fan09}).

During the polar field reversals the two hemispheres develop to
some extent independently (\opencite{sva13}); as a result the
polar fields complete their reversals not synchronously. There
exist time intervals when the global solar field loses its dipole
structure and behaves like a monopole (\opencite{wil72};
\opencite{kot09}). For example, during the  solar cycle 24 the
south polar field completed its reversal from positive to negative
in 2014. In contrast, the north polar field, after a series of
reversals changed its sign from negative in 2012 to positive at
the beginning of 2015 (\opencite{wang17}). As emphasized by
\inlinecite{kot09}, there is no theoretic explanation of the fact
that the positive or negative field can dominate practically on
the whole Sun for one year or more.

Reversal of  the polar magnetic field affects not only the
features of solar activity, but also considerably changes the
structure of interplanetary space. Significant difference was
found in the intensity-time profiles of the galactic cosmic rays
around the solar activity minima in the alternate solar magnetic
field polarities (\opencite{lock01}). One of manifestations of the
Sun's magnetic field asymmetry is the displacement of heliospheric
current sheet in the south direction (\opencite{mur03};
\opencite{erd10}). Observations of the interplanetary magnetic
field (IMF) have suggested a statistical tendency for the
heliospheric current sheet  to be shifted a few degrees southward
of the heliographic equator during the period 1965 -- 2010,
particularly in the years near sunspot minimum
(\opencite{wang11}).

The important information on the asymmetry of the Sun's magnetic
field can be obtained by consideration of the magnetic field net
flux and its changes in time. Net flux or flux imbalance  can be
defined as
\begin{equation}  \label{NetFl}
F_{net} = | F^{pos} | - |F^{neg}|.
\end{equation}
\textit{i.e.}, as a difference of positive and negative polarity
fluxes. The similar parameter was used by \inlinecite{tian03}
while considering magnetic fluxes for twelve solar active regions
in the Cycle 23. It was found that the fluxes of these regions
were appreciably imbalanced. For construction of the butterfly
diagram (\opencite{cho02}; \opencite{pet12}) signed net flux was
used also.

\inlinecite{cho02} found, using the magnetograms obtained
from the National Solar Observatory at Kitt Peak, that the maximum
and the median values of the flux imbalance for 137 active regions
were respectively 62\% and 9.5\%. The $10^\circ-40^\circ$ active
latitudinal zone in the individual hemispheres during the solar
maximum showed a flux imbalance of more than 20\%. This is reduced
to below 10\% when the entire Sun is considered.

When studying hemispheric asymmetry of magnetic fields
\inlinecite{pet15} explored correlations between three classes of
photospheric magnetic field: active region fields, which are
compact, intense, flux-balanced and usually bipolar in structure,
located equatorward of about $\pm 30^\circ$; polar fields, which
are located poleward of $\pm 60^\circ$ and are relatively weak,
almost unipolar and have large spatial scale, so that over most of
the solar cycle they appear dipole-like; and high-latitude surges
of field forming from decaying active regions, observed to stream
poleward between about $\pm 30^\circ$ and $\pm 60^\circ$ from
active latitudes in plume-like formations.

The present work continues the studies of \inlinecite{vernova14}
where the problem of polarity imbalance of magnetic flux for the
sunspot zone (latitudes form $-40^\circ$ to $40^\circ$ was
considered for the period 1976 -- 2003 and for fields greater than
$100$\,G. Here we used the data of SOLIS which enabled us to
continue the studied period till year 2016. Also, somewhat
different threshold values were used (fields greater than $50$\,G
and latitudes $5^\circ-40^\circ$). However, the main novelty of
the present work is that the polarity imbalance of the sunspot
zone is compared with that of the  high-latitude fields and their
connection with the quadrupole moment is established.

In papers \inlinecite{vernova16} and \inlinecite{vernova17} the
connection of magnetic field intensity with location at certain
heliospheric latitudes was studied. It was shown that in the
latitudinal distribution of the magnetic field averaged for three
cycles strong magnetic fields with strength of 15\,G and higher
are located in sunspot zone from $5^\circ$ to $40^\circ$. Strong
fields from both sides are surrounded by the weakest fields (less
than 5\,G: latitudes from $0^\circ$ to $5^\circ$ and from
$40^\circ$ to $60^\circ$). Above $40^\circ$ field strength does
not exceed 15\,G, except for a narrow strip of latitudes around
$70^\circ$ where polar faculae with fields from 15 to 50\,G are
observed.

In the present paper time-dependencies of the magnetic flux for
high-latitude and low-latitude  magnetic fields are studied as
well as the imbalance of positive and negative fluxes. Following
criteria for distinction between  low and high latitude regions
were used with this aim: low latitudes (sunspot zone) from
$5^\circ$ to $40^\circ$ in each hemisphere; high latitudes from
$40^\circ$ to $90^\circ$ in each hemisphere. The choice of
boundary latitude of $40^\circ$ allows to consider latitudinal
regions with quite different properties of magnetic fields. Only
strong fields ($B> 50$\,G) were considered in the low latitude
region, whereas in the high latitude region the weaker fields ($B
<50$\,G) were taken into account. The lower/upper limit of
magnetic field intensity ($B=50$\,G) was used to underline a role
of strong/weak fields in fluxes of low/high latitudes,
accordingly. The choice of this limit makes no major difference
and does not influence the main conclusions.

In Section~\ref{DataMeth}  we describe the data and discuss the
method applied in the article. Section~\ref{Time} is devoted to
the time changes of absolute values of magnetic fluxes. In
Section~\ref{Imbal} positive and negative magnetic fluxes and
their imbalance for a) high latitudes and b) sunspot zone are
considered separately. In Section~\ref{Build} we discuss and
interpret the obtained results and consider four building blocks
of the 22-year magnetic cycle. In Section~\ref{Concl} the main
conclusions are drawn.

\section{Data and Method}\label{DataMeth}
For our study we used synoptic maps of the photospheric magnetic
field  produced at the NSO Kitt Peak (from 1976 to 2016). Data for
1976 -- 1977 had many gaps and were not included in our analysis.
Combining data sets of two devices, KPVT for 1978 -- 2003
(available at \url{ftp://nispdata.nso.edu/kpvt/synoptic/mag/}) and
SOLIS for 2003 -- 2016 (available at
\url{https://magmap.nso.edu/solis/archive.html}) allowed to study
evolution of magnetic fields for nearly four solar cycles.
Synoptic maps have the following spatial resolution: $1^\circ$ in
longitude (360 steps), 180 equal steps in the sine of the latitude
from $-1$ (south pole) to $+1$ (north pole). Thus, every map
consists of $360\times180$ pixels of magnetic field strength
values in Gauss. On the assumption that the fields are radial to
the solar surface, the average line-of-sight component from the
original observations was divided by the cosine of the
heliocentric angle (the angle between the line-of-sight and the
vertical direction).

Noisy values near the poles
were detected and replaced by a cubic spline fit to valid values
in the polar regions. Although the instrumental noise per
resolution element is reduced by the map construction process, the
data are less reliable at the poles
(in the region with a viewing angle larger than $80^\circ$)
because of noisy measurements associated with the limb and the relatively
small number of original measurements that cover the poles.

However, in the sine of the latitude representation of
the Kitt Peak data, only one latitude zone corresponds to the
range $80^\circ - 90^\circ$, and so
this part of the data becomes negligible after the averaging
and does not affect the results of the paper.
The NSO data are complete in the
sense that where observational data are missing, such as when the
tilt of the Sun's axis to the ecliptic causes one of the polar
regions to be invisible, interpolated values are used (Durrant and
Wilson, 2003). Using as initial data magnetic field synoptic maps,
we have calculated fluxes for different intensities of magnetic
fields and for different latitudinal ranges.

\section{Time changes of magnetic fluxes}\label{Time}
\begin{figure}
\begin{center}
\includegraphics[width=0.75\textwidth]{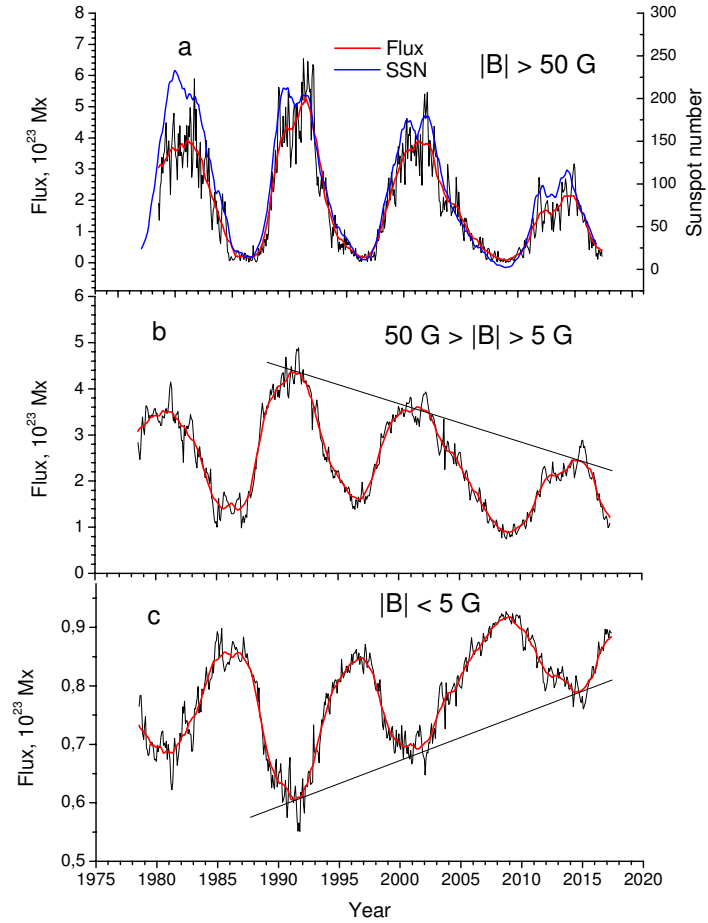}
\caption{Time changes of the magnetic flux for different ranges of
strength (absolute values) for the period 1978 -- 2016: (a) fields
with $|B|>50$\, G and sunspot number (SSN);
(b)~$50$\,G\,$>|B|>5$\,G; (c) $|B|<5$\, G. Black line -- magnetic
flux for each Carrington rotation; red line -- smoothed curves,
which were obtained using 20-rotation running mean.}
\label{compval}
\end{center}
\end{figure}

The magnetic field strength for both hemispheres (1978 -- 2016)
shows a nearly symmetric distribution of positive and negative
fields with 65.2\% of the pixels in the strength range $0 - 5$\,G,
whereas pixels with  the strength above 50\,G occupy only 2.6\% of
the solar surface. Magnetic fields in the $5 - 50$\,G strength range
occupy 32.2\%. Strong magnetic fields of both polarities occupy
a relatively small part of the Sun's surface. However just the strong
magnetic fields of active regions give the main contribution to
the total magnetic flux (see Figure~\ref{compval}).

Let us consider time changes of the magnetic flux for the whole
range of heliolatitudes, but for different values of magnetic
field strength. In Figure~\ref{compval}a the flux of the strongest
magnetic fields ($B> 50$\,G) is presented. Only absolute values of
magnetic field strength were considered.  For comparison 13-month
smoothed monthly sunspot number is shown
(\url{http://sidc.oma.be/silso/DATA/SN_ms_tot_V2.0.txt}) by blue
line in Figure~\ref{compval}a. The magnetic flux tracks the
11-year cycle based on sunspot number, reaching a maximum around
the time of the second (Gnevyshev) maximum of solar activity (SA).
In the minimum of SA the flux falls practically to zero. In three
consecutive maxima (Cycles 22, 23 and 24) the flux maximum
decreased monotonously more than two times. For the sunspot number
this decrease is observed during four solar cycles.

For magnetic fields from 5 to 50\,G (Figure~\ref{compval}b) time
changes closely agree with time changes of the strongest fields
(correlation coefficient $R=0.91$). Changes with the 11-year cycle
and monotonous decrease of the flux maxima can be seen for three
last solar cycles (noted in Figure~\ref{compval}b by a straight
line). The difference between Figures~\ref{compval}a and
\ref{compval}b is that the flux of $5-50$\,G fields is lower in
comparison with the flux of the strongest fields. Other difference
is in that the $5-50$\,G flux does not fall anywhere below a
threshold $~1\times10^{23}$\,Mx.

Very special time change display fields with strength $B < 5$\,G
which develop in antiphase with the solar cycle
(Figure~\ref{compval}c). In years when the flux of the strong
fields reaches its maximum, fields  of $B < 5$\,G have a minimum.
The minimum flux increases from Cycle 22 to 24 while for the
fields of higher strength the flux maximum falls (compare
Figures~\ref{compval}b and \ref{compval}c). The maximum values of
fluxes of weak magnetic fields ($B <5$\,G) are about ten times
lower, than the total flux of stronger fields ($B > 5$\,G).
Correlation coefficient between fields of $0-5$\,G and fields $B >
50$\,G is $R = - 0.91$, and between fields of $0-5$\,G and fields
of $5-50$\,G it is equal to $R = - 0.98$. Features of the
weakest-field change are in accordance with the results obtained
by analysis of SOHO/MDI data for 1996 -- 2011 which show that
magnetic structures with low fluxes change in antiphase with the
solar cycle (\opencite{jin}).

\begin{figure}[h]
\begin{center}
\includegraphics[width=0.75\textwidth]{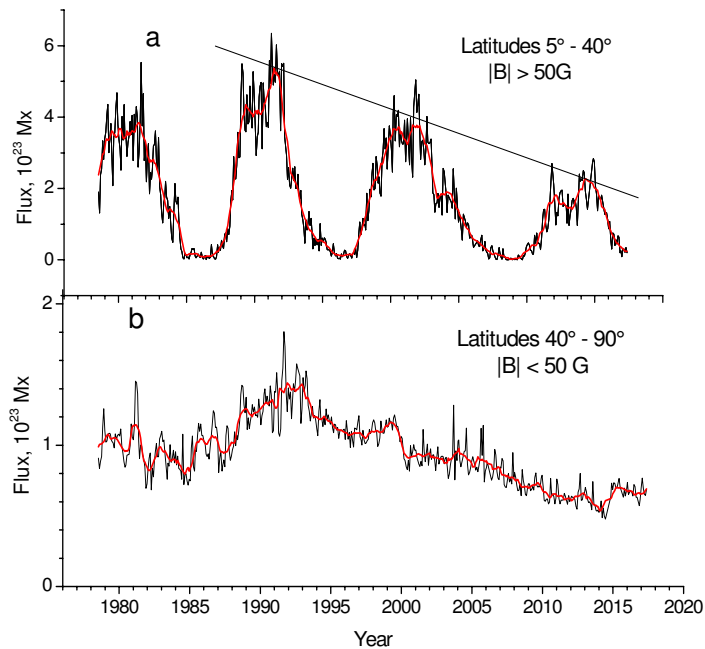}
\caption{Change of magnetic flux (absolute values) in different
latitude ranges for Solar Cycles 21--24: (a) low latitudes
$5^\circ - 40^\circ$, $B> 50$\,G; (b) high latitudes $40^\circ -
90^\circ$, $B <50$\,G. Black line -- magnetic flux for each
Carrington rotation; red line -- 20-point moving average.}
\label{complat}
\end{center}
\end{figure}

In Figure~\ref{complat} change of magnetic flux for Solar Cycles
21--24 is shown separately for low latitudes $5^\circ - 40^\circ$,
$B> 50$\,G (a) and for high latitudes $40^\circ - 90^\circ$, $B
<50$\,G (b). Only absolute values of magnetic field strength were
considered. For low latitudes the main flux is connected with
active regions; thus, the magnetic flux at low latitudes changes
with a 11-year cycle (Figure~\ref{complat}a), and the flux maximum
coincides with the time of the second maximum of Gnevyshev (in the
same way as in Figure~\ref{compval}a). The time course of
magnetic flux for high latitudes (Figure~\ref{complat}b) does not
show appreciable recurrence. The basic feature of magnetic flux at
high latitudes is monotonous decrease (approximately two times)
from a maximum around year 1991, to the minimum values around
2014, that is within 25 years. According to \inlinecite{obr09},
after 1980 the magnetic moment of the solar dipole showed a
tendency to gradually decrease and, in 2007, it has already
reached values lower than those at the beginning of the 20th
century. The polar field during the Cycle 23 minimum was about
40\% weaker than during two previous cycles (\opencite{wang09}).
One can see the general tendencies of magnetic flux decrease both
at high, and at low latitudes: falling of the polar field during
three cycles coincides with decline of the maxima of these cycles
for low latitude fields.

\section{Imbalance of positive and negative magnetic fluxes}\label{Imbal}
\subsection{High latitudes}\label{Highlat}
The results obtained previously (\opencite{vernova16};
\opencite{vernova17}) show that the latitude $40^\circ$  is the
boundary, above which the main contribution is made by the
magnetic fields lower than 50\,G.
\begin{figure}[h]
\begin{center}
\includegraphics[width=0.95\textwidth]{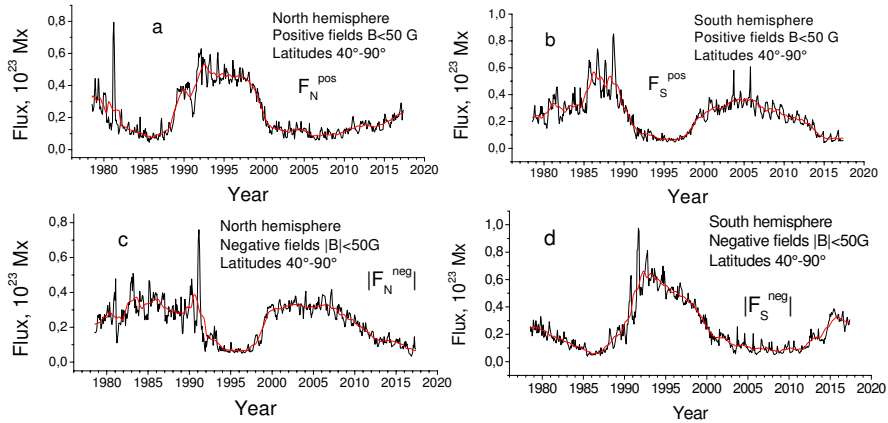}
\caption{Magnetic fluxes of high-latitude fields from $+40^\circ$
to $+90^\circ$ and from $-40^\circ$ to $- 90^\circ$ ($B <50$\,G):
(a),(b) -- positive fluxes of high latitudes of the northern
($F_{N}^{pos}$) and the southern ($F_{S}^{pos}$) hemispheres; (c),
(d)  -- absolute values of negative fluxes in the northern
($|F_{N}^{neg}|$) and the southern ($|F_{S}^{neg}|$) hemispheres.
Black lines -- flux values for each solar rotation, red lines --
20 points smoothing.} \label{highflux}
\end{center}
\end{figure}

At the same time, latitudes from $40^\circ$ to $90^\circ$ break
into three latitudinal strips, each of which corresponds to
different intensities of magnetic fields and different
manifestations of solar activity: 1) fields of $15-50$\,G from
$65^\circ$ to $75^\circ$ (polar faculae); 2) fields of $5-15$\,G
from $60^\circ$ to $90^\circ$ (coronal holes); 3) fields lower
than 5\,G from $40^\circ$ to $60^\circ$. Time changes and flux
imbalance   proved to be similar for these field groups which
allows to combine the three groups and to study magnetic fluxes of
high latitude fields from $+40^\circ$ to $+90^\circ$ and from $-
40^\circ$ to $- 90^\circ$ ($B <50$\,G). In Figures~\ref{highflux}a,b
positive fluxes of high latitudes of the northern ($F_{N}^{pos}$)
and the southern ($F_{S}^{pos}$) hemispheres are presented. For
negative fluxes in Figures~\ref{highflux}c,d their absolute values
($|F_{N}^{neg}|, |F_{S}^{neg}|$) are presented.

Positive and negative fluxes for each hemisphere change with the
period of 22 years opposite in phases to each other. Transitions
from low values to high ones and from high to low values occur
quickly, in comparison with wide flat maxima that is most
pronounced in the northern hemisphere. It is worth noting the
sharp increases of the flux (black line, without smoothing) in the
northern hemisphere lasting for several solar rotations in 1981.2
and in 1991.2. The peak in 1981.2 occurred in the positive flux,
whereas sharp  increase of negative flux produced the peak in
1991.2. Approximately at the same periods of time (in 1982 and
1991) \inlinecite{wang04}) found the highest peaks in the IMF strength
during last 40 years, and these peaks coincided with peaks in the
equatorial dipole strength.

\begin{figure}[h]
\begin{center}
\includegraphics[width=0.75\textwidth]{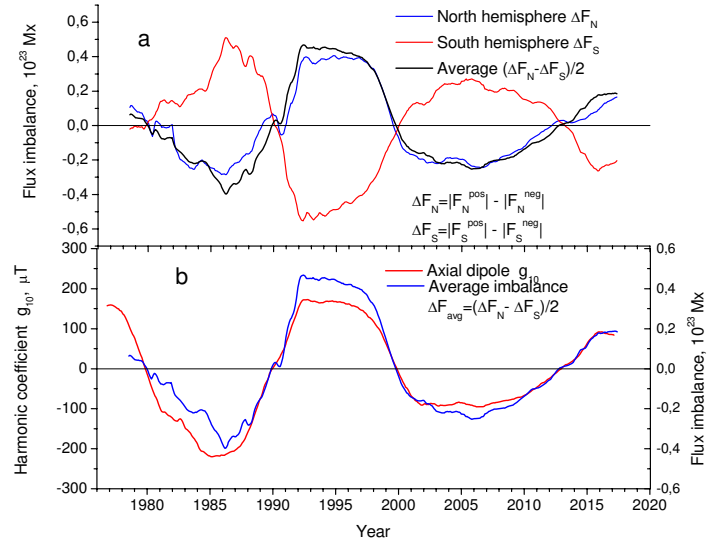}
\caption{(a) Imbalance of magnetic fluxes for the northern
($\Delta F_{N}=F_{N}^{pos} - |F_{N}^{neg}|$, blue curve) and the
southern ($\Delta F_{S}=F_{S}^{pos} - |F_{S}^{neg}|$, red curve)
hemispheres. The average imbalance ($\Delta F_{avg} = (\Delta
F_{N} - \Delta F_{S})/2$, black curve). (b) The dipole moment
$g_{10}$ of the multipole expansion of the photospheric magnetic
field (data of the WSO observatory).} \label{imbg10}
\end{center}
\end{figure}

In Figure~\ref{imbg10}a the imbalance of magnetic fluxes for the
northern ($\Delta F_{N}=F_{N}^{pos} - |F_{N}^{neg}|$,   blue
curve) and the southern ($\Delta F_{S}=F_{S}^{pos} -
|F_{S}^{neg}|$, red curve) hemispheres are shown. The average
imbalance ($\Delta F_{avg} = (\Delta F_{N} - \Delta F_{S})/2$,
black curve) changes in phase with the northern hemisphere
imbalance. For latitudes above $40^\circ$ dominating fields in
each hemisphere are those whose sign coincides with the sign of
the polar field in this hemisphere. This sign changes once in 11
years after polar field reversal. Thus, the imbalance of positive
and negative fluxes in a separate hemisphere changes with the
22-year period. The dipole moment $g_{10}$ of the multipole
expansion of the photospheric magnetic field according to
potential-field source-surface model (PFSS) (\opencite{ho86}) is
presented in Figure~\ref{imbg10}b (data of the WSO observatory,
available at \url{http://wso.stanford.edu/}). Comparison of the
average imbalance with the dipole moment $g_{10}$ shows their good
coincidence (Figure~\ref{imbg10}b). These results provide evidence
that the fields in the latitude range of $40^\circ-90^\circ$ are
directly connected with dipole component of  the Sun's magnetic
field.
\begin{figure}
\begin{center}
\includegraphics[width=0.75\textwidth]{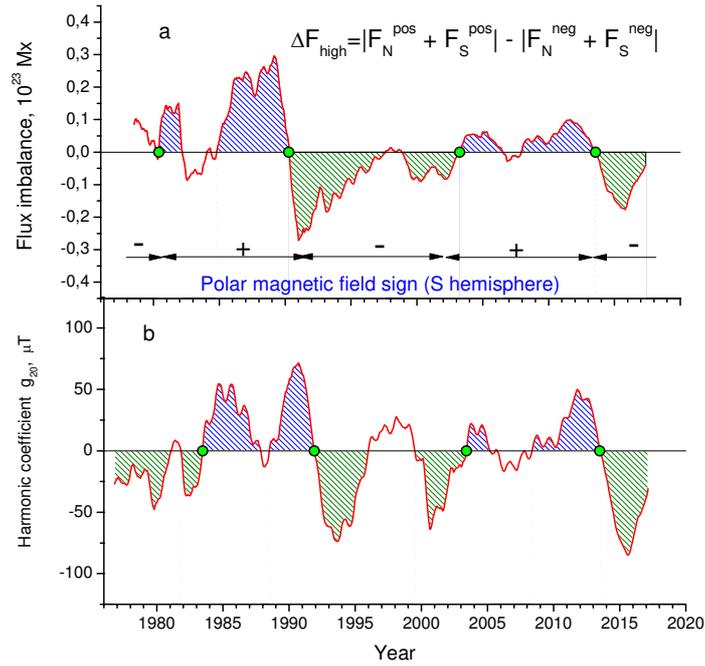}
\caption{(a) Imbalance $\Delta F_{high}$ of positive and negative fields of
both polar caps for the fields of $B <50$\,G
($\Delta F_{high} = |F_{N}^{pos} + F_{S}^{pos}| - |F_{N}^{neg} + F_{S}^{neg}|$).
(b) Axial quadrupole  moment $g_{20}$ of
the photospheric magnetic field.} \label{imbhigh}
\end{center}
\end{figure}

Imbalance $\Delta F_{high}$ of positive and negative fields of
both polar caps for the fields of $B <50$\,G was calculated
(Figure~\ref{imbhigh}a) as the sum of the signed high latitude
fluxes for the northern hemisphere $F_{N}^{pos}, F_{N}^{neg}$
(from $+40^\circ$ to $+90^\circ$ and for the southern one
$F_{S}^{pos}, F_{S}^{neg}$ (from $-40^\circ$ to $-90^\circ$):
\begin{equation}  \label{Imbal1}
 \Delta F_{high} = |F_{N}^{pos} + F_{S}^{pos}| - |F_{N}^{neg} + F_{S}^{neg}|
\end{equation}

Throughout four solar cycles the strict regularity of the
imbalance change is observed (change of the imbalance sign is
marked in Figure~\ref{imbhigh}a by green circles). From one
maximum of SA to another the imbalance looks like two peaks of one
sign divided by the period of the SA minimum. The sign of the
imbalance remains constant during 11 years from one polar field
reversal to the other. Thus the full period of change of the
imbalance sign makes 22 years. In the lower part of
Figure~\ref{imbhigh}a, the polarity of the polar magnetic field in
the southern hemisphere is displayed. Shading marks time intervals
when the imbalance sign coincides with the sign of polar field in
the southern hemisphere: positive sign -- dark blue shading, and
negative sign -- green shading. Most of the time the sign of the
imbalance coincides with the sign of the polar magnetic field in
the southern hemisphere.

In Figure~\ref{imbhigh}b the axial quadrupole  moment $g_{20}$ of
the photospheric magnetic field is shown (data of the WSO
observatory). Good coincidence of the imbalance sign with the sign
of the quadrupole moment can be seen. Not only the sign, but also
the basic features of the time course of the imbalance and the
quadrupole  moment are very close. The study of the north-south
asymmetry of the solar and heliospheric magnetic field
(\opencite{bra}) showed  that during activity minima the magnitude
of the northern magnetic field is smaller than the magnitude of
the southern magnetic field. According to the PFSS model the
dominant multipoles  are the dipole, the hexapole, and the
quadrupole. The sign of the quadrupole component is opposite to
that of the dipole and hexapole in the north, but the three
components have the same sign in the south which leads to
domination of  the southern hemisphere fields. This agrees with
the fact that the imbalance sign coincides always with the sign of
the polar field of the southern hemisphere
(Figure~\ref{imbhigh}à).

The obtained results allow us to draw the following conclusions
for high-latitude fields (from $40^\circ$ to $90^\circ$ in each
hemisphere), $B <50$\,G. The imbalance of positive and negative
fields in a separate hemisphere and the total imbalance of high-latitude
fields for two hemispheres change with the 22-year period. Change of the
sign of the imbalance occurs near reversal of the polar field. However,
while in a separate hemisphere the imbalance sign changes similarly to the
polar field in this hemisphere and to the dipole moment $g_{10}$, the
sign of the total imbalance for two hemispheres always coincides with
the sign of the polar field in the southern hemisphere. Change of the
sign of the total imbalance coincides with the sign change of the quadrupole
moment $g_{20}$.

\subsection{Active zone}\label{Actzone}
In our papers (\opencite{vernova16};  \opencite{vernova17}) it was
shown that strong magnetic fields which occupy latitude range from
$5 ^\circ$ to $40^\circ$ are surrounded from both sides by the
weakest fields ($B < 5$\,G, latitudes from $0^\circ$ to $5^\circ$
and from $40^\circ$ to $60^\circ$).

In the present paper for the study of strong fields in the sunspot
zone we set the following boundaries: field strength $B> 50$\,G
and latitudinal regions from $5^\circ$ to $40^\circ$. Positive and
negative magnetic fluxes of the near-equatorial region
$\pm5^\circ$ display  special features which we do not consider in
this paper.

\begin{figure}[t]
\begin{center}
\includegraphics[width=0.75\textwidth]{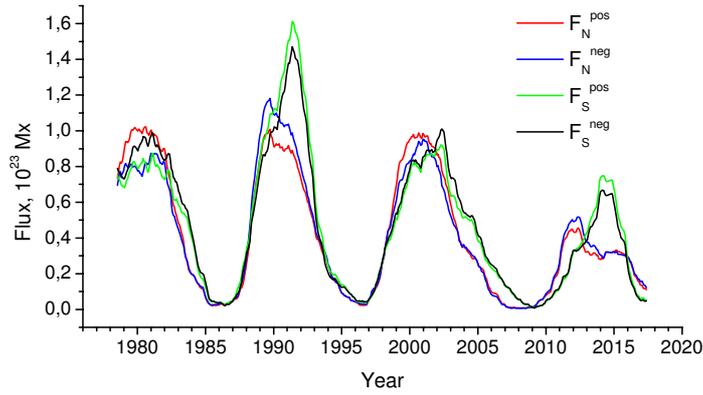}
\caption{Fluxes of strong magnetic fields ($B>50$\,G) for the
sunspot zone ($5^\circ - 40^\circ$): positive ($F_{N}^{pos}$,
$F_{S}^{pos}$) and negative ($F_{N}^{neg}$, $F_{S}^{neg}$) fluxes
(absolute values) for the northern and the southern hemispheres.}
\label{fourflux}
\end{center}
\end{figure}

In Figure~\ref{fourflux} fluxes of strong magnetic fields
($B>50$\,G) are presented for the sunspot zone ($5^\circ -
40^\circ$). For each synoptic map four different characteristics
of magnetic flux were obtained: absolute values of positive and
negative fluxes for the northern and the southern hemispheres --
$F_{N}^{pos}$, $F_{S}^{pos}$, $F_{N}^{neg}$, $F_{S}^{neg}$. All
these fluxes follow the 11-year cycle of solar activity. Some
difference between fluxes can be seen at periods of high solar
activity.  For each of solar cycles these four fluxes depending on
its polarity can be interpreted as the magnetic flux of the
leading or the following sunspots of one of the solar hemispheres.
The signs of the leading/following sunspots remain constant during
solar cycle from one minimum to the next one when sunspots of a
new cycle appear with polarities opposite to the previous ones.

\begin{figure}
\begin{center}
\includegraphics[width=0.75\textwidth]{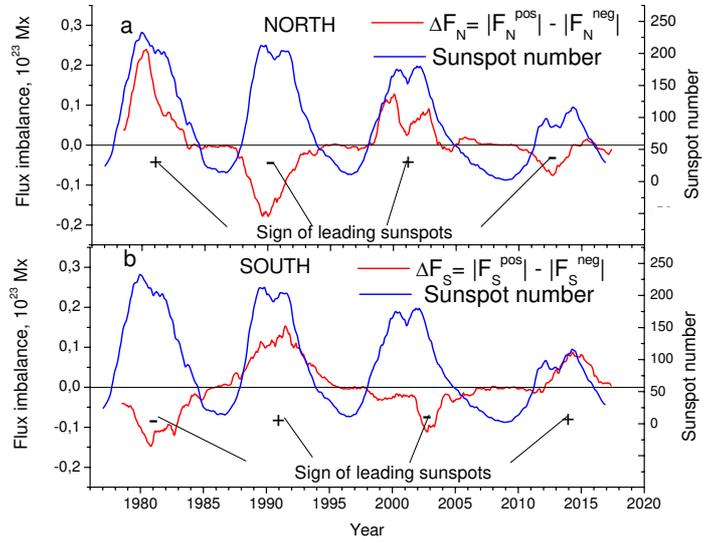}
\caption{(a) Flux imbalance between positive and negative fluxes
for the northern hemisphere $\Delta F_{N} = F_{N}^{pos} -
|F_{N}^{neg}|$. (b) Flux imbalance between positive and negative
fluxes for the southern hemisphere $\Delta F_{S} = F_{S}^{pos} -
|F_{S}^{neg}|$. Blue line denotes sunspot number. The sign of the
leading sunspots is shown for each hemisphere.} \label{imbhemi}
\end{center}
\end{figure}

Flux imbalance between positive and negative fluxes for the
northern hemisphere $\Delta F_{N} = F_{N}^{pos} - |F_{N}^{neg}|$
(Figure~\ref{imbhemi}a) varies with the 22-year cycle and reaches
extrema during maxima of solar activity. The flux imbalance passes
through zero around the minima of solar activity. Thus, from one
minimum to another, the sign of the difference between positive
and negative fluxes ($\Delta F_{N}$) does not change. In
Figure~\ref{imbhemi}a the sign of leading sunspots in bipolar
sunspot groups is shown. The sign of the flux imbalance coincides
with the sign of leading sunspots in the northern hemisphere.
Similar results are obtained for the southern hemisphere: the sign
of the imbalance $\Delta F_{S} = F_{S}^{pos} - |F_{S}^{neg}|$
always coincides with the sign of leading sunspots
(Figure~\ref{imbhemi}b). The imbalances of positive and negative
fluxes in each of the solar hemispheres show a 22-year recurrence
that is directly connected with the Hale cycle. Evidently, for
each of solar hemispheres magnetic flux of the leading sunspots
exceeds that of the following sunspots.

\begin{figure}
\begin{center}
\includegraphics[width=0.75\textwidth]{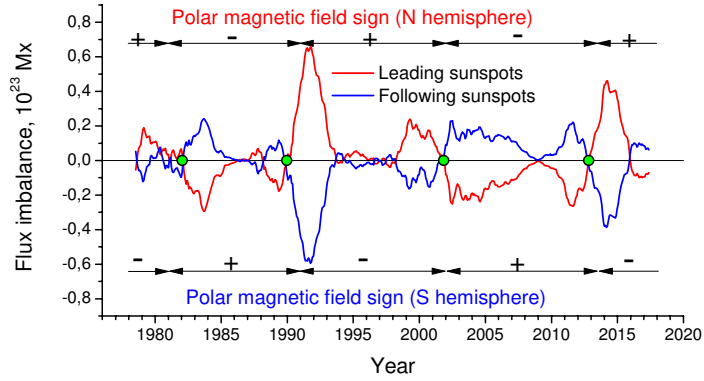}
\caption{Flux imbalance of leading sunspots of the northern and the southern
hemispheres $\Delta F_{lead}$ (red line) and flux
imbalance of following sunspots $\Delta F_{foll}$ (blue line).
Times of polar field reversal are marked by green
circles. The signs of the polar magnetic fields in the
northern and southern hemispheres are shown, respectively,
in the upper part and in the lower part of the figure.} \label{leadfol}
\end{center}
\end{figure}
There arises a question: if the fluxes of leading sunspots of two
hemispheres are compared, which of these  fluxes  will be higher?
It is similarly possible to consider the flux imbalance of
following sunspots of two hemispheres. For example, during Solar
Cycle 21, leading sunspots of the northern hemisphere  had
positive polarity (corresponding flux $F_{N}^{pos}$),  while the
leading sunspots of the southern hemisphere had negative  polarity
(corresponding flux $F_{S}^{neg}$). Then the imbalance of leading
sunspots can be defined as $\Delta F_{lead} = F_{N}^{pos} - |
F_{S}^{neg}|$. During the next solar cycle the leading sunspots in
both hemispheres change their polarity. In this case the imbalance
of leading sunspots will be $\Delta F_{lead} = F_{S}^{pos} - |
F_{N}^{neg}|$. Flux imbalance of leading sunspots of the northern
and the southern hemispheres $\Delta F_{lead}$ is presented in
Figure~\ref{leadfol} (red line). Flux imbalance of following
sunspots $\Delta F_{foll}$ (blue line) is in antiphase with the
imbalance of the leading sunspot fluxes with correlation
coefficient  $R= - 0.98$. At the time of polar field reversal both
parameters change their sign (marked by green circles). The sign
of the imbalance between fluxes of leading sunspots of two
hemispheres changes with a 22-year magnetic cycle in the same way
as the sign of the polar magnetic field in the northern hemisphere
(shown in the upper part of Figure~\ref{leadfol}). The imbalance
between fluxes of following sunspots repeats the sign of the polar
magnetic field in the southern hemisphere (shown in the lower part
of Figure~\ref{leadfol}).

It is possible to show that the imbalance of fluxes of leading
sunspots and the imbalance of fluxes of following sunspots define
the sign of north-south asymmetry of the magnetic field in the
activity zone. North-south asymmetry can be defined as:
\begin{equation}  \label{Asym}
 \Delta_{NS} = (F_{N}^{pos} + |F_{N}^{neg}|) - (F_{S}^{pos} +
 |F_{S}^{neg}|)
\end{equation}
The sign of the imbalance of leading sunspots is opposite to the
sign of the imbalance of following sunspots, hence, dominating
leading sunspots and dominating following sunspots will have
different signs. As fields of leading and following sunspots in
the same hemisphere have opposite signs, it follows from this that
both fluxes of leading sunspots of a hemisphere, and fluxes of
following sunspots of the same hemisphere will dominate
simultaneously over corresponding fluxes of the other hemisphere
(N-S asymmetry). The similar conclusion was made by
\inlinecite{pet12} for magnetic fields during decrease of Cycle 23
and ascent of Cycle 24.

Considering two phases of a 11-year cycle, from a minimum before
reversal and from reversal to a minimum, we will show that change
of domination of hemispheres (change of N-S asymmetry sign) occurs
during polar field reversal and during minimum of SA.
Figure~\ref{leadfol} shows that the sign of the imbalance of
leading sunspots always coincides with the sign of the polar field
in the northern hemisphere. It follows that those sunspots which
have the same sign, as the polar field of the northern hemisphere
will dominate. Thus, from the  minimum to the reversal the flux of
leading sunspots of the northern hemisphere which has the sign of
the polar field in this hemisphere, will always exceed the flux of
leading sunspots of the southern hemisphere (the northern
hemisphere dominates). After the reversal the sign of the polar
field in the northern hemisphere changes and coincides with the
sign of leading sunspots in the southern hemisphere (the southern
hemisphere dominates from the reversal to the minimum).

During SA minimum the domination returns to the northern
hemisphere  because leading and following sunspots in each of
hemispheres change their signs, whereas the sign of the polar
field remains unchanged. As a result north-south asymmetry changes
its sign both during polar field reversal and during minimum of
SA, so that  the northern hemisphere will always dominate from the
minimum to the reversal, but the southern hemisphere will dominate
from the reversal to the minimum (See Figure 2 in
\inlinecite{vernova14}).

\begin{figure}
\begin{center}
\includegraphics[width=0.75\textwidth]{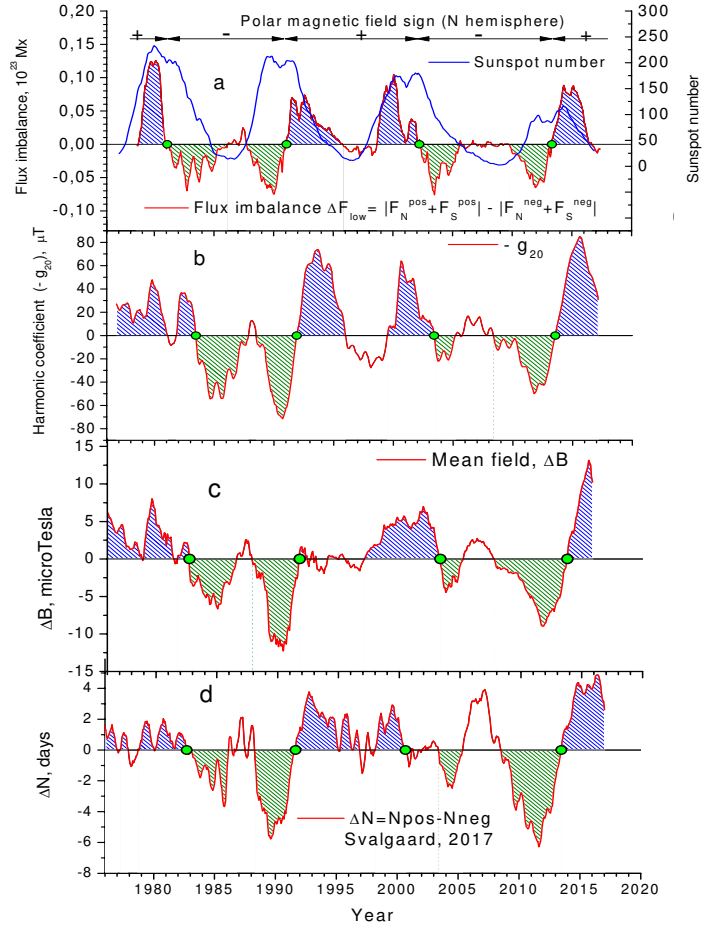}
\caption{(a) Total imbalance
$\Delta F_{low} = |F_{N}^{pos} + F_{S}^{pos}| - |F_{N}^{neg} + F_{S}^{neg}|$
of two sunspot zones (latitudes from $+5^\circ$ to
$+40^\circ$ and from $-5^\circ$ to $-40^\circ$).
Blue line -- sunspot number. The sign of the polar field in
the northern hemisphere is shown in the upper part of the figure.
(b) Reversed  quadrupole moment $-g_{20}$.
(c) Sun's mean magnetic field (SMMF) $\Delta B$.
(d) The difference between numbers of days with positive and negative
polarity of the interplanetary magnetic field (according to
L. Svalgaard).} \label{imblow}
\end{center}
\end{figure}

In the same way as the imbalance of magnetic field for two polar
caps  $\Delta F_{high}$ (Formula~(\ref{Imbal1})), one can define
the total imbalance of positive and negative fields for both
sunspot zones $\Delta F_{low}$ (latitudes from $+5^\circ$ to
$+40^\circ$ and from $-5^\circ$ to $-40^\circ$):
\begin{equation}  \label{Imbal2}
 \Delta F_{low} = |F_{N}^{pos} + F_{S}^{pos}| - |F_{N}^{neg} + F_{S}^{neg}|
\end{equation}
The total imbalance of two sunspot zones is presented in
Figure~\ref{imblow}a for magnetic fields $B > 50$\,G (red line).
The total imbalance changes in time similarly to the imbalance of
leading sunspots, but the imbalance of leading sunspots is about
five times higher. Change of the imbalance sign (marked in
Figure~\ref{imblow} by green circles) occurs during the period
close to the polar field reversal. The sign of the polar field in
the northern hemisphere is shown in the upper part of
Figure~\ref{imblow}a. The imbalance sign does not change within 11
years, thus the full period of the sign change makes 22 years. The
sign of the imbalance coincides with the sign of the polar field
in the northern hemisphere.  Time intervals of coincidence are
marked by shading for two polarities: positive (blue shading) or
negative (green shading).  As can be seen from comparison of the
flux imbalance with SA cycle (Figure~\ref{imblow}a, the dark blue
line) the imbalance changes with a strict regularity: each solar
cycle contains two parts -- one with a positive imbalance, another
with negative. In Figure~\ref{imblow}b the change of the
quadrupole moment taken with a reversed sign ($-g_{20}$) is shown
(the shading as in Figure~\ref{imblow}a). It is evident that the
sign of the magnetic flux imbalance and the sign of the quadrupole
moment ($-g_{20}$) change with the same periodicity.

To clarify the physical meaning of the imbalance one should
compare it with the north-south asymmetry $\Delta_{NS}$. In
Figure~\ref {leadfol} the change of the imbalances $\Delta
F_{lead}$ and $\Delta F_{foll}$ is shown. The north-south
asymmetry $\Delta_{NS}$ defined by Formula (\ref{Asym}) can be
rewritten as the difference between $\Delta F_{lead}$ and $\Delta
F_{foll}$, whereas the total imbalance for sunspot zone (see
Figure~\ref{imblow}a) is given by the sum of  $\Delta F_{lead}$
and $\Delta F_{foll}$. Since $\Delta F_{lead}$ and $\Delta
F_{foll}$ have opposite signs (see Figure~\ref {leadfol}), the
value of imbalance is significantly lower than that of the
north-south asymmetry.  Thus, the total imbalance provides a more
subtle characteristic of magnetic fields. All the more surprising
is that this characteristic has such regular structure and is so
closely related with 22-year and 11-year cycles.

It is of interest to compare the flux imbalance with the Sun's
mean magnetic field (SMMF). The SMMF represents an imbalance of
the magnetic flux ($\Delta B$) integrated over the entire visible
solar disk (\opencite{garc99}). It is small, less than $\pm 1$\,G,
reaching a typical value of about 0.15\,G during solar minimum.
SMMF is important in many problems of solar physics. Its influence
on the interplanetary magnetic field is strong, which can be
deduced from a high correlation between the two.  Daily values of
the mean magnetic field of the Sun (Sun-as-a-Star) in microTesla
(WSO data:
\url{http://wso.stanford.edu/meanfld/MF_timeseries.txt}) were
averaged over period of one month.  For comparison with the
imbalance of the sunspot-zone fluxes (Figure~\ref{imblow}a) the
monthly values of $\Delta B$  were smoothed over 20 points
(Figure~\ref{imblow}c). The values of SMMF differ from the
threshold of $B > 50$\,G chosen by us; however, the basic features
of the imbalance change can be observed in the mean field of the
Sun. The sign of  the SMMF also displays the 22-year periodicity
and changes around the time of the polar field reversal.

The question arises, whether the imbalance of positive and
negative fields is a purely local effect or it is reflected in the
structure of the heliosphere? To clarify this point, data  on the
polarity of the IMF (data of L. Svalgaard, see:
\url{http://www.leif.org/research/spolar.txt}) were used. The
difference between numbers of days with positive and negative
polarity of the IMF was calculated (Figure~\ref{imblow}d). For
convenience of comparison with the flux imbalance
(Figure~\ref{imblow}a) coincidence of polarities is marked by
shading. Good agreement between the imbalance and the  IMF
polarity can be seen, except for years 2005 -- 2008 of the
prolonged minimum of solar activity. The sign of the difference
between positive and negative days of the IMF changes with the
22-year period and coincides mainly with the sign of the polar
field in the northern hemisphere. Thus, the imbalance of the
magnetic flux of the Sun can be the cause of the asymmetry
observed in the IMF.

Results obtained for fields of the sunspot zone (from $5^\circ$ to $40^\circ$
 in each hemisphere), $B>50$\,G testify:
\begin{enumerate}
\item[1)]  The imbalance in a separate hemisphere changes with the
22-year cycle and the imbalance sign changes near the solar
activity minimum. The imbalance sign coincides with the sign of
leading sunspots. \item[2)]  The sign of the total imbalance for
two hemispheres also changes with the 22-year cycle, but the sign
change occurs near the reversal of the polar magnetic field. The
imbalance sign always coincides with the sign of the polar field
in the northern hemisphere. Total imbalance of positive and
negative fluxes shows similar evolution as the quadrupole
component $(-g_{20})$ of the photospheric magnetic field.
\end{enumerate}

\section{Building blocks of a 22-year cycle}\label{Build}
\begin{figure}
\begin{center}
\includegraphics[width=0.75\textwidth]{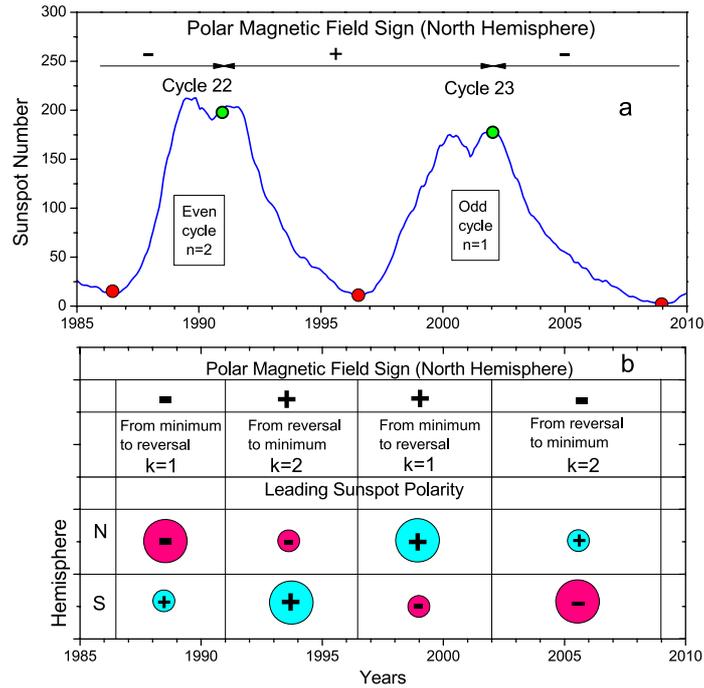}
\caption{Polar magnetic field sign and leading sunspot polarity
for each of the building blocks in Cycles 22 and 23.}
\label{scheme}
\end{center}
\end{figure}

Different manifestations of the magnetic-flux  imbalance
considered above, have one common feature: they appear regularly
during the period of  four solar cycles. We will show that this
regularity can be expressed by several simple formulas. Solar
dynamo models establish the connection of the 11-year cycle of
solar activity to the 22-year cycle of magnetic polarity. The
scheme of the polarity change of  local and global fields is
presented in Figure~\ref{scheme}. Two solar cycles are shown in
Figure~\ref{scheme}a: the even cycle and the following odd cycle.
Such choice is connected with the results of Ohl who found a good
correlation between SA of two consecutive cycles: an even and the
following odd cycle. On the other hand correlation between an odd
cycle and the following even one is very weak. According to the
Ohl scheme, an even cycle with the succeeding odd cycle form a
pair, a single whole (\opencite{nag09}; \opencite{pon02}). We use
number $n$ to denote parity of the cycle. Thus, parity of the
solar cycle $n = 1$ corresponds to an odd solar cycle, $n = 2$ to
an even one.

In the course of a 22-year magnetic cycle there are moments when
the relation between the polarities of the global and local
magnetic fields changes. These moments are related either to the
change of the Sun's global magnetic-field polarity during high
solar activity (polar magnetic-field reversal, marked in
Figure~\ref{scheme}a by green circles), or to the alternation of
the polarities of leading and following sunspots at the solar
activity minimum (minima of SA are marked by red circles). In the
minimum of SA there is a change of the leading-sunspot sign in
each of hemispheres. During the transition from an even to an odd
cycle leading sunspots of a new cycle in the northern hemisphere
will have the positive sign which coincides with the sign of the
polar field in this hemisphere. During the transition from an odd
cycle to an even the sign of leading sunspots in the northern
hemisphere also will coincide with the sign of the polar field,
however it will be a negative sign.

Another situation will take place at the reversal of the polar
magnetic field (around maximum of SA). After reversal and up to SA
minimum the sign of the polar field in the northern hemisphere
will coincide with the sign of leading sunspots in the southern
hemisphere. The following is remarkable: in both cases,
\textit{i.e.}, throughout all magnetic cycle, leading sunspots of
that hemisphere whose sign coincides with the polar-field sign of
the northern hemisphere (Figure~\ref{scheme}b) will dominate in
the magnetic flux. Thus, a 22-year cycle is composed of four
intervals (we will call them ``building blocks''): during each of
these intervals the relation between polarities  of the polar
magnetic field and of the leading sunspots does not change. These
four intervals are from the solar-activity minimum to the magnetic
field reversal of the Sun and from the reversal to the next
minimum in an even 11-year solar cycle, and analogously for the
next odd cycle. The relations between the polarities of the global
and local magnetic fields will repeat themselves in the next
22-year cycle. Connection of global and local magnetic field
polarities in a 22-year magnetic cycle is illustrated  by
Figure~\ref{scheme}b. Each solar cycle can be divided in two parts
according to the phase of the cycle: $k = 1$ corresponds to the
interval of the 11-year cycle from the minimum up to the reversal;
$k = 2$ to the interval from the reversal up to the minimum. The
sign of the polar magnetic field in the northern hemisphere is
shown at the top of the Figure~\ref{scheme}a and in the scheme of
Figure~\ref{scheme}b for each of the four quarters of the 22-year
magnetic cycle.

The leading sunspots in each of hemispheres are displayed as
circles with corresponding polarity. Large circles show dominating
leading sunspots. It can be seen that in every building block the
sign of the dominating leading sunspots coincides with the sign of
the polar magnetic field in the northern hemisphere. The scheme of
Figure~\ref{scheme}b shows that at each boundary of two
characteristic intervals domination of leading sunspots passes
from one hemisphere to the other one. In the first quarter of the
22-cycle the northern hemisphere dominates, in the next quarter
the southern hemisphere dominates and so on. As is shown in
Figure~\ref{leadfol}, fluxes of leading and following sunspots in
the same hemisphere will dominate simultaneously over
corresponding fluxes of leading and following sunspots of the
other hemisphere. Thus, from the minimum to the reversal the
northern hemisphere always dominates, whereas  from the reversal
to the minimum -- the southern hemisphere dominates (N-S
asymmetry).

Following simple formulas describe the magnetic field polarities
for different phases ($k=1, 2$) of the odd ($n=1$) and even
($n=2$) solar cycles:

1. The sign of the polar magnetic field of the northern hemisphere
$F_N$;  the sign of leading-sunspot flux imbalance  $\Delta F_{lead}$; the
sign of positive and negative flux imbalance for sunspot zones
$\Delta F_{low}$ (latitudes from $+5^\circ$ to $+40^\circ$ and from $-5^\circ$ to
$-40^\circ$) are determined by two factors: the parity of the
solar cycle $n$, and the phase of the 11-year cycle $k$:
\begin{equation}  \label{Form1}
\left. \begin{array}{c}
{\rm sign}\, F_{N}  \\
{\rm sign}\, \Delta F_{lead} \\
{\rm sign}\, \Delta F_{low}
\end{array}
\right\}  = (-1)^{n+k}.
\end{equation}

2. The sign of the polar magnetic field of the southern hemisphere
$F_S$; the sign of  following-sunspot flux imbalance  $\Delta
F_{foll}$; the sign of positive and negative flux imbalance for
high latitudes from  $+40^\circ$ to  $+90^\circ$ and from
$-40^\circ$ to $-90^\circ$ $\Delta F_{high}$ are determined by the
same two factors $n$, $k$:
\begin{equation}  \label{Form2}
\left. \begin{array}{c}
{\rm sign}\, F_{S}  \\
{\rm sign}\, \Delta F_{foll} \\
{\rm sign}\, \Delta F_{high}
\end{array}
\right\} = (-1)^{n+k+1}.
\end{equation}

3. The sign of leading sunspots  in the northern hemisphere; the
sign of imbalance between positive and negative fluxes in the
northern hemisphere  $\Delta F_{N}$ are determined by the parity of
the solar cycle n:
\begin{equation}  \label{Form3}
\left. \begin{array}{c}
{\rm sign}\, (lead.\, sunspot) \\
{\rm sign}\,  \Delta F_{N}
\end{array}
\right\} = (-1)^{n+1}.
\end{equation}
4. The sign of leading sunspots  in the southern hemisphere; the
sign of imbalance between positive and negative fluxes in the
southern hemisphere  $\Delta F_{S}$ are determined by the parity of
the solar cycle n:
\begin{equation}  \label{Form4}
\left. \begin{array}{c}
{\rm sign}\, (lead.\, sunspot) \\
{\rm sign}\, \Delta F_{S}
\end{array}
\right\} =  (-1)^{n}.
\end{equation}
5. The sign of the north-south asymmetry depends on the phase of
the 11-year cycle  k (before or after the reversal):
\begin{equation}  \label{Form5}
{\rm sign}\,  \Delta_{NS} = (-1)^{k+1}
\end{equation}

6. Previously we considered the longitudinal distribution of
sunspots (\opencite{vernova04}) and of photospheric magnetic
fields (\opencite{vernova07}). The longitudinal
distribution for the ascending phase and the maximum sharply
differ from the longitudinal distribution for the descending phase
and the minimum of the solar cycle. Active longitudes change by
$180^\circ$ when SA evolves from the ascending phase to the
descending one. The maximum of the longitudinal distribution is
attained at $180^\circ$ for the ascending phase and the maximum of
the solar cycle when the polarities of the leading sunspot and of
the global field coincide in each hemisphere, and at
$0^\circ/360^\circ$ for the descending phase and the minimum when
these polarities are opposite. The active longitude is determined
by the phase of solar cycle $k$:
\begin{equation}  \label{Form6}
Active\,\, Longitude =  \pi k
\end{equation}

The above formulas evidence that four building blocks of the
22-year magnetic cycle manifest itself in periodic changes of
magnetic field polarities. We believe that representation of the
22-year magnetic cycle as consisting of four characteristic
intervals (building blocks) can be useful for studying the
processes underlying the observed changes of magnetic field
polarities. This representation does not merely establish the fact
of alternation of global and local field polarities, but also
states the change of domination of certain magnetic field group
that is reflected in changes of solar activity and
solar-terrestrial relations. In fact, many observations show that
behavior of the solar activity manifestations changes
significantly both for different phases of the solar cycle and for
the odd and even cycles.

Some examples of this connection are presented below. Difference
between phases of ascent and descent leads to occurrence of a
hysteresis for many indexes of SA (\opencite{bach}). The
hysteresis effect that shows up as a nonunique relationship among
the emissions from the photosphere, chromosphere, and corona
during the rising and declining phases of solar and stellar
activity was analyzed by \inlinecite{bru16}. Solar-terrestrial
relations also display the dependence on the phase of the cycle.
It was shown that ionospheric indices fo2 and Ap weakly depend on
the level of solar activity, but the effect of hysteresis is
clearly seen (\opencite{bru16a}). The number of M-class and
X-class flares tends to follow the sunspot number, but there is a
tendency to have more flares on the declining phase of a sunspot
cycle (\opencite{hath}).

The 22-year periodicity of  SA displays itself as the difference
between two  successive  maxima of the 11-year cycle, which
follow  the Gnevyshev--Ohl rule: the maximum of an even cycle is
lower than the maximum of the following odd cycle. Studying the
Fraunhofer lines in the solar spectrum variations
\inlinecite{liv07} found that solar minimum around 1985 was clearly seen
in the  high photosphere lines, but the following minimum in 1996
was missing, perhaps indicating a role for the 22-year Hale cycle.

Cosmic-ray modulation shows strong dependence both on the 11-year
cycle of SA, and on the 22-year magnetic cycle. When the solar
magnetic field in the northern hemisphere is positive, the
galactic cosmic rays drift downward from the solar poles toward
the Earth and then out along the warped current sheet to the
boundary of the heliosphere (\opencite{lock01}). When it is
negative, the drift pattern is the opposite. The difference in the
appearance of the two halves of the 22-year solar magnetic cycle,
peaked and flat-topped, supports the role of drifts near the
cosmic ray intensity maxima. The intensity of cosmic rays at the
solar activity minima also depends on the solar magnetic field
polarity. \inlinecite{sing05} observed differences between time
lags of the solar activity and cosmic ray intensity in odd and
even cycles as well as differences in the shape, size etc. of
hysteresis loops during odd and even cycles. The time lag between
cosmic ray intensity and the solar index is different in odd
($10-14$ months) and even ($1-4$ months) cycles. Differences in
time lag between periods of $A < 0$ polarity ($9-10$ months) and
$A > 0$ polarity ($3-5$ months) was also found ($A>0$ corresponds
to the positive polarity of the polar field in the north
hemisphere). These examples show the influence of different phases
of the 22-year magnetic cycle on solar activity and heliosphere.

\section{Conclusions}\label{Concl}

In this paper we studied polarity imbalance of photospheric
magnetic field for high latitudes (from $40^\circ$ to $90^\circ$
in each hemisphere) and for sunspot zone (from $5^\circ$ to
$40^\circ$ in each hemisphere) during four solar cycles (Solar
Cycles 21--24). We used the threshold $B <50$\,G for high
latitudes and $B >50$\,G for the sunspot zone. For these two
latitude zones we considered imbalance of positive and negative
magnetic fields for each hemisphere and for both hemispheres
together. For sunspot zone we calculated polarity imbalance for
leading sunspots of two hemispheres as well as for following
sunspots. All these polarity imbalances display regular structure
and change with the 22-year period.

The threshold $40^\circ$ separates two latitude intervals whose
imbalances ($\Delta F_{high}$ and $\Delta F_{low}$) develop in
antiphase. The sign of the total imbalance for high latitudes
(from $40^\circ$ to $90^\circ$ and from $-90^\circ$ to
$-40^\circ$) coincides with the sign of the polar field in the
southern hemisphere and also with the imbalance of following
sunspots according to Formula (\ref{Form2}). On the other hand,
the sign of the imbalance for sunspot zone (from $5^\circ$ to
$40^\circ$ and from $-40^\circ$ to $-5^\circ$) coincides with the
sign of the polar field in the northern hemisphere as well as with
the imbalance of leading sunspots (see Formula (\ref{Form1})). We
obtained a good agreement of the imbalance of high-latitude fields
with the quadrupole moment $g_{20}$, while for low-latitude fields
there is an agreement with $-g_{20}$, the quadrupole moment taken
with the opposite sign.

It was noted in \inlinecite{mur16} and \inlinecite{wang14} that
the quadrupole component plays in important role in the shift of
the heliospheric current sheet southward. Our paper reveals a very
close relation of the magnetic-field imbalance with the quadrupole
moment. It looks reasonable to suggest that just the imbalance of
sunspot-zone fields, which always shows domination of the fields
with the same sign as the polar field in the northern hemisphere,
results in the southward shift of the heliospheric current sheet.

Imbalances for high latitudes and for sunspot zone have opposite
signs and, as a result, the imbalance for all latitudes has lower
values. However, while magnetic flux for sunspot zone is
approximately three times greater than magnetic flux for high
latitudes (Figure~\ref{complat}), the total imbalance for sunspot
zone is about two times less than the total imbalance for high
latitudes (see Figure~\ref{imbhigh}a and Figure~\ref{imblow}a).
This difference between the imbalances for sunspot zone and for
high latitudes (which is especially  clearly seen in Cycle 22)
shows that total imbalance for the whole range of latitudes cannot
be zero.

In the course of a 22-year magnetic cycle there are four intervals
(from the solar-activity minimum to the magnetic-field reversal of
the Sun and from the reversal to the next minimum in an even
11-year solar cycle, and analogously for the next odd cycle) when
the relation  between polarities  of the polar magnetic field and
of the leading sunspots does not change. We believe that this
representation of the 22-year magnetic cycle as consisting of the
four characteristic intervals (``building blocks'') can be useful
in the study of the processes on the Sun and of solar-terrestrial
relations.

The obtained results show that in addition to well-known 22-year
periodic patterns in the change of global and local fields there
also exists a 22-year periodicity in the change of imbalance
of both high-latitude and low-latitude fields.


%
 \begin{acks}
NSO/Kitt Peak data used here are produced cooperatively by
    NSF/NOAO, NASA/GSFC, and NOAA/SEL.
This work utilizes SOLIS data obtained by the NSO Integrated
Synoptic Program (NISP), managed by the National Solar
Observatory, which is operated by the Association of Universities
for Research in Astronomy (AURA), Inc. under a cooperative
agreement with the National Science Foundation. Wilcox Solar
Observatory data used in this study was obtained via the web site
http://wso.stanford.edu courtesy of J.T. Hoeksema. We are thankful
to Prof. L. Svalgard for providing data on the IMF sector
structure polarity.
 \end{acks}


%
%
%

\end{article}
\end{document}